\documentclass[twocolumn,showpacs,amsmath,amssymb]{revtex4}
\usepackage{graphicx}
\usepackage{dcolumn}
\usepackage{bm}

\newcommand{\ba}{\begin{eqnarray}}
\newcommand{\ea}{\end{eqnarray}}
\newcommand{\ban}{\begin{eqnarray*}}
\newcommand{\ean}{\end{eqnarray*}}
\newcommand{\be}{\begin{equation}}
\newcommand{\ee}{\end{equation}}
\newcommand{\bd}{\begin{displaymath}}
\newcommand{\ed}{\end{displaymath}}

\newcommand{\n}[1]{\label{#1}}
\newcommand{\nn}{\nonumber}
\newcommand{\Eq}[1]{(\ref{#1})}

\newcommand{\bBox}{\bar{\phantom{!}\Box\phantom{!}}}

\newcommand{\x}{{\mathbf x}}
\newcommand{\y}{{\mathbf y}}
\newcommand{\U}{{\cal U}}
\newcommand{\J}{{\cal J}}
\newcommand{\E}{{\cal E}}

\newcommand{\pa}{\partial}

\newcommand{\hh}{\, ,\hspace{0.5cm}}
\newcommand{\hhh}{\, ,\hspace{0.2cm}}

\begin{document}

\draft


\title{Relativistic gyratons in asymptotically AdS spacetime}
\author{Valeri P. Frolov$^{1,3}$}
\email{frolov@phys.ualberta.ca}
\author{Andrei Zelnikov$^{1,2,3}$}
\email{zelnikov@phys.ualberta.ca}
\affiliation{$^{1}$Theoretical Physics Institute, University of Alberta,
  Edmonton, AB, Canada, T6G 2J1\\
  $^{2}$Lebedev Physics Institute,
  Leninsky prospect 53, 119991, Moscow Russia\\
  $^{3}$Asia Pacific Center for Theoretical Physics,
  Pohang 790-784, Korea}
\date{\today}

\begin{abstract}
We study the gravitational field of a spinning radiation beam-pulse
(a gyraton) in a $D-$dimensional asymptotically AdS spacetime. It is
shown that the Einstein equations for such a system reduce to a set
of two linear equations in a $(D-2)-$dimensional space. By
solving these equations we obtain a metric which is  an exact
solution of gravitational equations with the (negative) cosmological
constant. The explicit metrics for $4D$ and $5D$ gyratons in
asymptotically AdS spacetime are given and their properties are
discussed.
\end{abstract}

\pacs{04.70.Bw, 04.50.+h, 04.20.Jb \hfill Alberta-Thy-11-05}

\maketitle

\section{Introduction}

The gravitational field created by beams of radiation, and pulses of light has
been studied intensively since a pioneer paper by Tolman \cite{To} who found
the solution of gravitational equations in linear approximation.  The exact
solutions of the Einstein equations for the pencil of light has been found by
Peres \cite{Pe1,Pe2} and Bonnor \cite{Bo}.  The gravitational field of a
spinning beam-pulse of finite duration, a gyraton, generalizes these solutions
to the case when the beam-pulse carries an angular momentum
\cite{FrFu:05,FroIsrZel:05}. A typical example of a gyraton would be a pulse of
a circular polarized light or a modulated beam of ultrarelativistic particles
with a spin. The gravitational field of the gyraton is parametrized by a number
of arbitrary functions of the retarded time $u$. These functions arise through
the dependence on $u$ of the coefficients in mode expansion of the
gravitational field. They describe profiles of the energy density and angular
momenta distributions of the gyraton propagating in an asymptotically flat
$D$-dimensional spacetime. The gyraton solutions in asymptotically flat
spacetimes belong to a general class of pp-waves. In the limit of an
infinitesimally short impulse and zero angular momentum the solutions describes
a gravitational field of an ultrarelativistic particle - a gravitational shock
wave \cite{AiSe}.

In this paper we generalize results for gyratons in asymptotically flat
spacetime \cite{FrFu:05,FroIsrZel:05} to the case when a spacetime is
asymptotically AdS. That is, we obtain exact solutions for the geometry of the
gyraton propagating in an asymptotically AdS background. For zero angular
momentum these solutions belong to the type of Siklos spacetimes \cite{Siklos}
generalized to higher dimensions. In the limit of a $\delta(u)$-like impulse
these solutions correspond to gravitational shock waves in AdS spacetime.
Similar to shock waves in a flat spacetime the solutions can be derived using
an infinite boost of the gravitational field of a point particle provided the
energy of the particle is kept fixed  
\cite{DrayTHooft:85,Ferrari:88,Barrabes:01} (the Penrose limit). Shock wave
metrics can also be obtained from a global AdS space by cut-and-paste technique
\cite{DrayTHooft:85,Sfetsos:95,Kaloper:05}. In the string theory gravitational
shock waves propagating in a flat background got much attention since they do
not receive $\alpha'$ corrections \cite{HorowitzSteif:90,AmatiKlimcik:89}. This
property can be generalized also to the case of gravitational shock waves
propagating in AdS metrics \cite{HorowitzItzhaki:99}. The proof of this
property is based on geometrical arguments, namely, on the fact that all scalar
invariants constructed from the Riemann tensor and its derivatives are the same
for pure AdS and for AdS shock wave solutions. In the present paper we prove
that this   geometrical property is also valid in more general case of the
gyraton AdS spacetime.

The paper is organized as follows. Section \ref{Section2} collects formulas for
the gyraton metric in the asymptotically flat spacetime obtained earlier in
\cite{FrFu:05,FroIsrZel:05}. This is done in order to fix notations. These
formulas are also used later when we are discussing the asymptotically flat
space limit of the obtained gyraton metrics in the AdS spacetime. In Section
\ref{Section3} a set of equations for a gyratons in the asymptotically AdS
spacetime is derived. A general solution of these equations is obtained in
Section \ref{Section4}. Explicit metrics for gyratons in $4D$ and $5D$
asymptotically AdS spacetimes are obtained in the Sections \ref{Section5} and
\ref{Section6}, respectively. Some properties of these solutions and their
applications and generalizations are discussed in Section \ref{Section7}.
Appendices contain expressions for the scalar and vector Green functions in the
AdS spacetime, which are used in the main text.


\section{Gyratons in asymptotically flat spacetime}\label{Section2}

A gravitational field of a gyraton propagating in an asymptotically flat
spacetime has been found recently in \cite{FrFu:05,FroIsrZel:05}. It is
described by the metric
\ba\n{flat.gyraton}
ds^2=-2\,du\,dv+d{\bf x}^2+\Phi~du^2
+2~({\bf A},d{\bf x})~du\, ,\\ \nn
u=x^1,\hskip 0.5cm v=x^2, \hskip 0.5cm{\bf x}=x^a.
\ea 
The functions $\Phi(u,{\bf x})$ and ${\bf A}(u,{\bf x})$ do not depend on $v$.
The spatial part of the metric \Eq{flat.gyraton} in the $(D-2)-$dimensional
hyperplane transverse to the direction of the motion of the gyraton is flat,
\be
\n{flat} d{\bf x}^2=\delta_{ab}dx^adx^b=\sum_{a=3}^{D} (dx^a)^2\, . 
\ee
$l=l^{\mu}\pa_{\mu}=\pa_v$ is a null Killing vector. The metrics of this form
are the most general D-dimensional null Brinkmann metrics \cite{Brinkmann:25}
with flat transverse space, which sometimes are called $pp$-wave metrics
\cite{Tseytlin:95}.  In what follows we assume that Latin indices $a,b,\dots$
for the coordinates in the transverse plane run from 3 to $D$. We assume that
the sum is taken over the repeated Latin indices and omit the summation symbol.
We denote covariant derivatives with respect to the flat spatial metric
$\delta_{ab}$  by colon, $()_{:a}$.

The functions $\Phi$ and $A_a$ can be considered as a scalar and a vector field
in the $(D-2)-$dimensional Euclidean space which depend also on an external
parameter $u$. The metric \Eq{flat.gyraton} is invariant under  the coordinate
transformation
\ban
v\rightarrow v+\lambda(u,x^a)\, ,
\ean
provided the functions $\Phi$ and $A_a$ transform as follows
\be\n{gauge}
A_a\to A_a-\lambda_{,a}\hhh
\Phi\to \Phi-2\lambda_{,u}\, .
\ee
We shall also use the following notation
\be\n{Fab}
F_{ab}=\pa_aA_b-\pa_bA_a
\ee
for the antisymmetric tensor in the $(D-2)-$plane. This tensor is
evidently invariant under the transformation \Eq{gauge}.

The nonzero components of the Ricci
tensor for the metric \Eq{flat.gyraton} are
\cite{FroIsrZel:05}
\ban
R_{ua}&=&{1\over 2}F_{ab}{}^{:b} ,\\
R_{uu}&=&-{1\over 2}\Phi^{:a}_{:a}+{1\over 4}F_{ab}F^{ab}+\partial_u
(A_a{}^{:a})\, .
\ean
Thus the Einstein equations reduce to the following two sets of equations in
$(D-2)-$dimensional flat space
\ba\n{flat.maxwell}
&&F_{ab}{}^{:b}=J_a,\\
\n{flat.scalar}
&&\Phi^{:a}_{:a}=-J+{1\over 2}F_{ab}F^{ab}+2\partial_u (A_a{}^{:a})\, ,
\ea
where
\be\n{J}
J_a=\kappa T_{ua}, \hskip 0.7cm J=\kappa [T_{uu}-{1\over D-2}\,g_{uu}T].
\ee
Here $\kappa=16\pi G$ and $G$ is the $D-$dimensional gravitational
coupling constant.

The first set of equations \Eq{flat.maxwell} formally coincides with the
Euclidean Maxwell equations in $D-2$-dimensional Euclidean space
(magnetostatics), $J_a$ playing the role of the current. The second equation
\Eq{flat.scalar} is similar to the equation for the electric potential with the
only difference that besides the charge distribution $J$ it contains an extra
source proportional to ${\bf F}^2$.

To solve the equations \Eq{flat.maxwell}-\Eq{flat.scalar} one should first find a
vector potential from the linear equation \Eq{flat.maxwell},  and
then substitute the obtained solution to the right hand-side of
\Eq{flat.scalar}. It is convenient to split $\Phi$ into two parts,
the first part being a solution for the source $J$, and the second one
being a solution for the distributed "charge" $J_F=-{1\over 2}{\bf F}^2$
\ba\n{Phi} 
&&\Phi=\varphi+\psi,\\ \n{Psi}
&&\varphi_{:a}^{:a}-2\partial_u (A_a{}^{:a})=-J,\hskip 0.5cm
\psi_{:a}^{:a}={1\over 2}F_{ab}F^{ab}.
\ea

Note that the combination
$
\Phi_{:a}^{:a}-2\partial_u(A_a{}^{:a})
$
remains invariant under the gauge transformation \Eq{gauge}.
Therefore, the equations \Eq{flat.maxwell},
\Eq{flat.scalar},\Eq{Phi}, and \Eq{Psi} are gauge invariant.

The source terms $J$ and $J_a$ vanish outside the position of the gyraton.
Solutions obtained in \cite{FrFu:05,FroIsrZel:05} describe the gravitational
field of the gyraton in the limit when its transverse size goes to zero. Though
it is possible to find an exact solution of the Einstein equations for an
arbitrary source of finite size, it makes sense to consider first point-like
distributions in the transverse space. Let us emphasized that in the general
case the solution (\ref{Psi}) is only formal and may not have a well-defined
sense. The reason is that for a point-like current, $F_{ab}$ has a singularity
at ${\bf x}=0$. If one considers this singular function as a distribution, one
needs to define what is the meaning of $F_{ab}F^{ab}$ in (\ref{Psi}). This
problem does not exist for a distributed source (gyraton).  At some small
distance outside the point-like source the vacuum solution of the Einstein
equations provides a description of the problem in question.  At the position
of the source this vacuum solution can be stitched to the metric created by
gyraton with a generic distribution of the energy density and angular momentum
( see a discussion of the problem at the end of Chapter II of the paper
\cite{FroIsrZel:05} ). In our analysis of the gyratons in the asymptotically
AdS spacetime we assume that the source terms are locally the same as in the
asymptotically flat case. One can reformulate this condition by requiring that
the local properties of the gravitational field near ${\bf x}=0$ are the same
in both cases.


\section{Gyratons in asymptotically AdS spacetime}\label{Section3}

Now consider a gyraton propagating  in the D-dimensional asymptotically
AdS background.

It is well known that a pure AdS spacetime is conformal to the
Minkowski one. Let us choose one of the spatial coordinates, say
$x^3\equiv z$, and consider the following metric
\be\n{AdS}
d\bar{s}^2=\bar{g}_{\mu\nu}dx^\mu dx^\nu
={L^2\over z^2}\left[-2~du~dv+d{\bf x}^2\right].
\ee
It is easy to show that this metric has constant curvature and obey
the Einstein equations
\be
\bar{R}_{\mu\nu}-{1\over 2}\bar{R}\bar{g}_{\mu\nu}=-\Lambda
\bar{g}_{\mu\nu}\, ,
\ee
where $\Lambda=-{(D-1)(D-2)/(2 L^2)}$. The constant $L$ in
(\ref{AdS})  is the radius of the curvature of the AdS world.
It is worthwhile to mention, that in the metric (\ref{AdS}) the
coordinate $z$ play a special role, and  by using the conformal
factor depending on $z$ one might expect that the symmetry of the
metric (\ref{AdS}) would be less than the symmetry of the
original flat metric. It does not happen. The spacetime \Eq{AdS} remains
homogeneous and isotopic, but instead of the Poincare group its
isometry group is $SO(D,1)$. In the solution (\ref{AdS}) the point
$z=0$ corresponds to the spatial infinity and $z=\infty$ is the horizon
defined for the set of observers sitting at rest at constant $z$.

As the ansatz for the gyraton metric in the asymptotically AdS
spacetime we use the following expression
\ba\nn
ds^2&=&{L^2\over z^2}\left[\right.-\left.2dudv+d{\bf x}^2\right.\\ \n{AdS.gyraton}
&+&\left.\Phi(u,{\bf x})du^2
+2({\bf A}(u,{\bf x}),d{\bf x})du\right].
\ea
Here again $z=x^3$ is one of the spatial coordinates. In the absence of the
gyraton $\Phi={\bf A}=0$ and the metric reduces to the pure AdS metric. This
property would be preserved asymptotically if one assumes that both functions
$\Phi(u,{\bf x})$ and ${\bf A}(u,{\bf x})$ vanish at the infinity of the
transverse space. In what follows we assume that this condition is satisfied.
Note also that in the limit $L\rightarrow\infty$, while the location of the
gyraton is kept near $z_0\rightarrow L$  and
$[(z-z_0)^2+(\x-\x_0)^2]/L^2\rightarrow 0$, one gets the asymptotically flat
gyraton geometry.

   Before discussing solutions of the Einstein equations of the form
\Eq{AdS.gyraton} note that this metric  has the following property: All local
scalar invariants constructed from the metric, the Riemann tensor and its
covariant derivatives are exactly the same as those for the pure AdS spacetime
\Eq{AdS}. This property is valid off-shell, i.e., the metric does not need to
be a solution of the Einstein equations. The proof of this statement is given
in Appendix \ref{AppendixC}.  This is a generalization to the case of
relativistic gyratons of the statement by Horowitz and Itzhaki
\cite{HorowitzItzhaki:99} which has been given in application to gravitational
shock waves in AdS.


Now let us return to the dynamics of the gyraton spacetime. Substituting the
ansatz \Eq{AdS.gyraton} into the Einstein equations
\ban
R_{\alpha\beta}-{1\over 2}R g_{\alpha\beta}+\Lambda g_{\alpha\beta}=8\pi
T_{\alpha\beta},
\ean
and using the same notations \Eq{flat},\Eq{Fab}, and \Eq{J} as in
asymptotically flat case, we obtain again two nontrivial equations
\ba\n{maxwell}
F_{ab}{}^{:b}&-&{D-2\over z} F_{az}=J_a,
\\ \n{scalar}
\Phi_{:a}^{:a} - {1\over 2}F_{ab}F^{ab} &-& 2\partial_u\left(A_a^{:a}\right)
\\&-&{D-2\over z}\left(\partial_z\Phi-2\partial_u A_z\right)=-J\nn .
\ea
Here $()_{:a}$ denotes the covariant derivatives with respect to the flat
metric in the transverse space.


\section{Solving the Einstein equations}\label{Section4}

\subsection{Magnetostatics in AdS space}

The analogy of \Eq{maxwell} with the "magnetostatic" Maxwell equations
\Eq{flat.maxwell} can be
made more precise if we rewrite \Eq{maxwell} as the Maxwell equations in a
fiducial N-dimensional Euclidean AdS space \Eq{EuclideanAdS}.
To do this let us introduce a fiducial Euclidean AdS metric $\tilde{g}_{AB}$
\ba\n{ADS}
\tilde{g}_{AB}={L^2\over z^2}\delta_{AB},\hskip 1cm
A,B=(3,\dots,N+2).
\ea
For simplicity we put the radius parameter $L=1$. It is not difficult to
restore proper dimensionality later. Let us denote the covariant derivatives in AdS
metric \Eq{ADS} by semicolon and covariant derivatives in flat
N-dimensional metric by a vertical bar. Then one has
\ban
F_{AB}{}^{;B}&=&
z^2\left(F_{AB}{}^{|B}
-{N-4\over z}F_{A z}\right).
\ean
One can see that the equation \Eq{maxwell} is identical to the Maxwell
equations in $N=D+2$ dimensional Euclidean AdS space and reads
\ba\n{ADSvector}
F_{AB}{}^{;B}&=&\tilde{J}_A, \hskip 1cm
\tilde{J}_A=z^2J_A,
\ea
\ba
F_{AB}=\pa_A A_B-\pa_B A_A,
\ea
\ba
A_A=(A_a,0,0,0,0), \hskip 1cm J_A=(J_a,0,0,0,0).
\ea
We call this space fiducial. The current $J_A$ and the vector potential $A_A$
depend only on $D-2$ coordinates $x^a$ and the retarded tine  $u$. To obtain a
solution for $A_A$ one needs to know the Green function for vector field in
Euclidean AdS space
\ba
A_A(x)=\int
d^Nx'\sqrt{\tilde{g}'}~G_{AB'}(x,x')~\tilde{J}^{B'}(x').
\ea
The vector Green function $G_{AB'}$ in the pure AdS space is well known
\cite{AllenJacob:1986,HFMMR:1999}. We present the explicit expression for it
in Appendix \ref{AppendixB}.


\subsection{Massless scalar field in AdS space}

Similarly to the vector field case we represent \Eq{scalar} in the form of the
scalar field equation in the fiducial higher dimensional Euclidean AdS space
\Eq{EuclideanAdS}.
\ba\n{gAB}
\bar{g}_{AB}={L^2\over z^2}\delta_{AB}, \hskip 0.5cm
A,B=(3,\dots,D+1,D+2).
\ea
One has $A_A=(A_a,0,0)$,
\ban
&&\Phi_{;A}^{;A}=
z^2\left(\Phi_{|A}^{|A}-{D-2\over z}\Phi_{,z}\right)
=z^2\left(\Phi_{:a}^{:a}-{D-2\over z}\Phi_{,z}\right),\\
&&A_A^{;A}=
z^2\left(A_A^{|A}-{D-2\over z}A_z\right)
=z^2\left(A_a^{:a}-{D-2\over z}A_z\right),\\
&&F_{AB}F^{AB}=z^4 F_{ab}F^{ab}.
\ean
For the scalar equations the dimensionality of the fiducial space $N=D$ is
chosen so that these equations reproduce \Eq{scalar}. Thus we can rewrite
equation \Eq{scalar} for the metric of the gyraton as an equation for the
functions $\Phi$ defined in the pure Euclidean AdS geometry with the proper
number of dimensions $(N=D)$. The scalar equation then reads 
\ba\n{ADSscalar}
\Phi_{;A}^{;A}=-z^2J+{1\over 2 z^2}
F_{AB}F^{AB}+2\partial_u \left(A_A^{;A}\right).
\ea
A solution of this problem expressed in terms of the scalar Green function 
(\ref{G2k}) in
the D-dimensional Euclidean AdS space is
\ba\nn
\Phi&=&\varphi+\psi,\\
\n{phi}
\varphi&=&\int d^Dx\sqrt{\bar{g'}}~G(x,x')~\bar{J}(x'),\\
\n{psi}
\psi&=&\int d^Dx\sqrt{\bar{g'}}~G(x,x')~\bar{J}_F(x'),\\ \nn
\bar{J}&=&\left[z^2 J -2\partial_u \left(A^{;A}_A\right)
\right],\hskip 0.5cm
\bar{J}_F=-{1\over 2 z^2} F_{AB}F^{AB}.
\ea


\section{$4D$ Gyratons}\label{Section5}

Let us consider a special case of a gyraton in $4D$ AdS. Its metric has the
form
\ban
ds^2={-2dudv+dz^2+dx^2+\Phi du^2+2(A_zdz+A_xdx)du\over z^2}.
\ean
The spatial components $J_a$ of the vector current are assumed to be localized
at $x=x_0$ and to have the same structure as in asymptotically flat case
\cite{FrFu:05,FroIsrZel:05}
\ba\nn
J_a={\kappa\over 2}j(u)z^2\epsilon_{ab}\partial_b[\delta(z-z_0)\delta(x-x_0)].
\ea
Here $\epsilon_{ab}$ is the 2D Levi-Civita symbol and the total angular
momentum  of the source is given by $\J=\int du\,j(u)$. The current satisfies
the conservation law
\ban
\partial^a J_a-{2\over z}J_z=0.
\ean
We look for the vector potential in the form
\ba\n{Aa}
A_a=z^2\epsilon_{ab}\partial_b\sigma(z,x),
\ea
which, evidently, satisfies the gauge condition
\ban
\partial^aA_a-{2\over z}A_z=0.
\ean
The two-dimensional field strength is given by the relation
\ba\n{F}
F_{ab}=\pa_a A_b-\pa_b A_a = -\epsilon_{ab}\pa_c(z^2\pa^c\sigma).
\ea
We restrict ourselves to the case when $F_{ab}$ decreases at spatial infinity,
then from the equations \Eq{maxwell} one gets
\ba\n{sigm}
{1\over z^2}\pa_a\left(z^2\pa^a
\sigma\right)=-{\kappa\over 2}j(u)\delta(z-z_0)\delta(x-x_0).
\ea
Looking at \Eq{F} and \Eq{sigm} one can see that in four dimensions due to the
Maxwell equations $F_{ab}$,  vanishes everywhere outside the location of the
gyraton
\footnote{Generically
one could add to $F_{ab}$ the solution of the homogeneous equation $\sim
z^2\epsilon_{ab}$ , but this term would violate our asymptotic requirements},
while $A_{a}$ is nontrivial.

The solution for $\sigma$ can be easily found. It reads
\ba\nn
\sigma&=&{\kappa\over 8\pi}j(u){1\over zz_0}\ln{\U\over\U+2}\\ \n{sigma}
&=&{\kappa\over 8\pi}j(u){1\over zz_0}\ln{(z-z_0)^2+(x-x_0)^2\over
(z+z_0)^2+(x-x_0)^2},
\ea
where
\ba\n{U}
\U={(z-z_0)^2+(x-x_0)^2\over 2z z_0}=2{r_{-}^2\over r_{+}^2-r_{-}^2}
\ea
and
\ba\n{r}
r_{\pm}=\sqrt{(z\pm z_0)^2+(x-x_0)^2}.
\ea
Thus we obtain
\ba\nn
A_a&=&{\kappa\over 8\pi}j(u)~z^2\epsilon_{ab}\partial_b \left[{1\over zz_0}
\ln{(z-z_0)^2+(x-x_0)^2\over (z+z_0)^2+(x-x_0)^2}\right]\\ \n{Aa}
&=&{\kappa\over \pi}j(u)~z^2\epsilon_{ab}\partial_b \left[{1\over
r_{+}^2-r_{-}^2}
\ln{r_{-}\over r_{+}}\right].
\ea

Now consider  the scalar equation \Eq{scalar} 
\ba\n{FF}
\partial^a\partial_a\Phi-{2\over z}\partial_z\Phi&=&-J+{1\over 2}F_{ab}F^{ab}.
\ea
where
\ban
J=-\kappa\sqrt{2}~\varepsilon(u)z^2~\delta(z-z_0)\delta(x-x_0).
\ean
A solution for $\Phi$ can be found by using the method discussed in the
previous section. In fact we use \Eq{phi} by introducing two extra dimensions
$\y=(y^1,y^2)$ and using scalar Green functions \Eq{scalarG} given in Appendix
\ref{AppendixA}. For $F_{ab}=0$ one has 
\footnote{For an extended gyraton $F_{ab}$ does not vanish in a region occupied
by the gyraton, so that ${\bf F}^2$-term in \Eq{FF} gives a non-trivial
contribution to $\Phi$. This contribution depends on the size of the gyraton as
well as on its structure. In the simplest case in the limit of a small size
gyratons this contribution results in the renormalization of the energy density
$\varepsilon(u)$. For details see \cite{FroIsrZel:05}. 
}
\ban
\Phi&=&-\kappa\sqrt{2}~\varepsilon(u)~\int
d\y~G_4(z,x,\y;z_0,x_0,0)\\ &=&-{\kappa\sqrt{2}\over
4\pi}\varepsilon(u)~zz_0~\left[(\U+1)\ln{\U\over \U+2}+2\right]\\
&=&-{\kappa\sqrt{2}\over8\pi}\varepsilon(u)~\left[(r_{+}^2+r_{-}^2)\ln{r_{-}\over
r_{+}}+r_{+}^2-r_{-}^2\right].
\ean
Here functions $\U$ and $r_\pm$ are defined by \Eq{U} and \Eq{r}.

Note that the four dimensional gyraton is a special case, because the 
equations for $F_{ab}$ assume that it vanishes outside the location of the
gyraton. Therefore, locally the vector function $A_a$ can be put to zero using 
a proper coordinate transformation \Eq{gauge}. In this gauge the metric
acquires the Siklos \cite{Siklos} form. However, one can not make $A_a$ to be
zero globally, since the gauge invariant contour integral $\oint A_a(u,\x)dx^a$
around the position of the gyraton is proportional to the angular momentum
density $j(u)$ which is non-zero. This is why we prefer to present the vector
potential $A_a$  in the form \Eq{Aa}, which is analogous to the vector
potential of the Bohm-Aharonov flux. Let us emphasize that, for higher
dimensions $D\ge 5$ the tensor $F_{ab}$ is non-trivial anyway.


\section{$5D$ Gyratons}\label{Section6}

In five dimensions AdS gyraton metric reads
\ba\nn
ds^2={-2dudv+(dx^a)^2+\Phi du^2+2A_adx^adu\over z^2},
\ea
where $x^a=(z,x^3,x^4)$.
The gyraton current has a form
\ba\nn
J_a={\kappa\over 2}z^3j(u)~\epsilon_{abc}n^c\partial_b[\delta^3(x^a-x^a_0)].
\ea
It is parametrized by a unit 3-vector $n^c$ which points in the
direction of the magnetic dipole. Without loss of generality one can
write
\ba
n^c=\delta^c_z\cos{\eta}+\delta^c_3\sin{\eta}, \hskip 1cm
\eta={\mathrm{const}}.
\ea
This current satisfies the conservation law
\ban
J_a^{:a}-{3\over z}J_z=0.
\ean

Let us consider the simplest case when the current is in the $(x^3,x^4)$
plane, i.e., dipole moment is directed along $z$-axis,  $n^c=(1,0,0)$. Then
\ba
J_a={\kappa\over 2} z^3
j(u)~\epsilon_{abz}\partial_b[\delta^3(x^a-x^a_0)].
\ea
In the Lorentz gauge
\ban
A_a^{:a}-{3\over z}A_z=0,
\ean
the vector potential can be written in the form
\ba
A_a=\epsilon_{abz}\partial_b[\sigma(x^c)].
\ea
The corresponding Maxwell tensor takes the form
\ban
&&F_{ab}=\epsilon_{abc} H^c, \hskip 1cm  H^c={1\over 2}\epsilon^{cef}F_{ef},\\
&&H^c=-\delta^c_z\pa_b\pa^b\sigma+\pa_z\pa^c\sigma,\\
&&F_{ab}=-\epsilon_{abz}\pa^c\pa_c\sigma+\epsilon_{abc}\pa_z\pa^c\sigma,\\
&&F_{ab}{}^{:b}-{3\over z}F_{az}
=-\epsilon_{abz}\pa^b\left[\pa_c\pa^c\sigma-{3\over
z}\pa_z\sigma\right].
\ean
From this we obtain the following equation for $\sigma$
\ba\n{sigma5}
\pa_c\pa^c\sigma-{3\over z}\pa_z\sigma=-{\kappa\over 2} z^3
j(u)\delta^3(x^a-x^a_0),
\ea
which coincides with the scalar field equation in $5D$ AdS spacetime.
Using the scalar Green functions \Eq{scalarG} we get
\ba\nn
\sigma&=&-{\kappa\over 4\pi}j(u)\,zz_0 {\left[\U+1-\sqrt{\U(\U+2)}\right]^2\over
2\sqrt{\U(\U+2)}}\\
&=&-{\kappa\over 64\pi}j(u){(r_{+}-r_{-})^4\over r_{+}r_{-}}.
\ea
Here
\ban
\U={(z-z_0)^2+\x^2\over 2zz_0},\hskip 0.5cm 
r_{\pm}=\sqrt{(z\pm z_0)^2+\x^2},
\ean
and $\x^2=(x^3-x^3_0)^2+(x^4-x^4_0)^2$.

In the vicinity of the current, i.e, when $r_{-}\rightarrow 0$ one
gets
\ba
\sigma\rightarrow -{\kappa\over 8\pi}j(u){z_0^3\over r_{-}}.
\ea

Contrarily to  4D case, the Maxwell tensor $F_{ab}$ in five dimensions does not
vanish outside the source. The corresponding "magnetic" field $H^c$ is
\ba\n{Hc}
H^c=\left(\pa^c-{3\over z}\delta^c_z\right)\pa_z\sigma+{\kappa\over
2} j(u)\delta^c_zz^3\delta(z-z_0)\delta(\x).
\ea
The equation for $\Phi$ involves the current $J$ (see \Eq{J}) and
\ban
J_F=-{1\over 2}F_{ab}F^{ab}=-H^cH_c.
\ean
The local terms, originating from $\delta$-function, have meaning
of proper energy of the gyraton and can be combined into the
redefined energy parameter $\E$ of the source. So, the nontrivial
contribution of angular momenta to the $\Phi$ component of the
gyraton metric comes from the first term in \Eq{Hc}.

The solution for the $g_{uu}$ component of the metric follows the same lines as
that of 4D case
\ban
\Phi&=&\varphi+\psi,\\
\varphi_{:a}^{:a}&-&{3\over z}\partial_z\varphi=-J,\\
\psi_{:a}^{:a}&-&{3\over z}\partial_z\psi=-J_F.
\ean
The scalar current is
\ban
J=-\kappa\sqrt{2}z^3\varepsilon(u)~\delta(x^a-x^a_0).
\ean
Because the equation for $\varphi$ is exactly of the same type as \Eq{sigma5},
we can write down the answer without further calculations
\ba\nn
\varphi&=&{\kappa\sqrt{2}\over2\pi}\varepsilon(u)zz_0 {\left[\U+1-\sqrt{\U(\U+2)}\right]^2\over
2\sqrt{\U(\U+2)}}\\
&=&{\kappa\sqrt{2}\over32\pi}\varepsilon(u){(r_{+}-r_{-})^4\over r_{+}r_{-}}.
\ea
Using the same Green function \Eq{scalarG} we can write the solution for $\psi$
\ba
\psi=\int dz'd\x' {\mathbf G}(z,\x;z'\x'){1\over z'^3} J'_F.
\ea
Here
\ban
{\mathbf G}(z,\x;z'\x')=-{1\over
8\pi\sqrt{((z-z')^2+\Delta\x^2)((z+z')^2+\Delta\x^2)}}\\
\times\left[z^2+z'^2+\Delta\x^2-\sqrt{((z-z')^2
+\Delta\x^2)((z+z')^2+\x^2)}\right]^2,
\ean
\ba\nn
\Delta\x^2=(\x-\x')^2=(x^3-x'^3)^2+(x^4-x'^4)^2,
\ea
and $J'_F=-H'^cH'_c$ is a function of $(z',z_0,\x')$ and $H_c$ is
defined by \Eq{Hc}.

\section{Summary and Discussions}\label{Section7}

The main result of this paper is the generalization of the gyraton
solutions \cite{FrFu:05,FroIsrZel:05} to asymptotically AdS
spacetimes. These metrics describe the gravitational field of
ultrarelativistic beam pulses with non-zero angular momentum
propagating in the AdS spacetime. We demonstrate how the method
proposed in \cite{FrFu:05,FroIsrZel:05} can be generalized to solve
the Einstein equations for the spacetime which is asymptotically AdS.
The corresponding solutions contain a number of arbitrary functions
of $u$ describing distributions of the energy density and angular
momenta of the beam pulse. As special examples we discuss 
the gyraton AdS metrics in 4 and 5 dimensions in detail.
In the absence of angular momentum the 4D AdS-gyraton solutions
reduce to the Siklos spacetimes \cite{Siklos}. In the absence of
rotation and for a $\delta$-function profile of the energy density,
the obtained solutions coincide with AdS shock waves metric
\cite{Podolsky:98}. 

The gyraton geometry in asymptotically flat spacetimes has the property that
all scalar invariants constructed from the curvature and its covariant
derivatives are zero \cite{FroIsrZel:05}. It is this property, that in the case
of shock waves made it possible to conclude that quantum and $\alpha'$
corrections to the metric are zero \cite{HorowitzSteif:90,AmatiKlimcik:89}. We
demonstrate that a similar property is valid for the gyraton AdS metrics.
Namely, all scalar invariants constructed from the Riemann tensor and its
covariant derivatives are the same both for the asymptotically AdS gyraton
metric and for exact AdS spacetime. It was shown earlier \cite{Kallosh:98} that
in the string theory $\alpha'$ corrections do not modify AdS solution. Thus,
according to the geometrical approach arguments \cite{HorowitzItzhaki:99} these
corrections, probably, should not modify the asymptotically AdS gyraton metric
as well.

It should be emphasized that we focused on the solutions outside the
region occupied by a gyraton. To obtain a total solution one needs to
solve the interior problem inside the region occupied by the gyraton
and to glue this solution with an exterior metric. Solutions for 4D
and 5D AdS gyratons presented in the paper (which formally has a
singularity at ${\bf x}=0$) generalize special solutions discussed in
\cite{FrFu:05,FroIsrZel:05}. 

The obtained AdS gyraton solutions can be used for study of the mini
black hole production in the collision of two ultrarelativistic
particles with spin moving in the AdS space.  It would be also
interesting to analyze these solutions in relation with the AdS-CFT
correspondence. In particular, a   gyraton may have a complex
structure which is encoded in its gravitational field.    According
to the AdS-CFT correspondence the asymptotic of this metric at the
AdS space infinity must be sufficient to obtain the complete
information about the gyraton structure. It might be possible since
all the multipole moments of the field in the AdS do not fall of
faster at infinity \cite{HorowitzHubeny:00}. It is interesting to discuss
this mechanism in more details.

\noindent
\section*{Acknowledgments}
\noindent
The authors are grateful to Werner Israel for stimulating discussions
and remarks. This work was supported by the Natural Sciences and
Engineering Research Council of Canada, by the Killam Trust and in
part by APCTP.


\appendix


\section{Curvature invariants}\label{AppendixC}
\subsection{Curvature}

Let us demonstrate that the metric \Eq{AdS.gyraton} has the
following property: All local invariants constructed from the metric,
the Riemann tensor and   its covariant derivatives are exactly the
same as those for the pure AdS spacetime \Eq{AdS}. This property is
valid off-shell, i.e., the metric does not need to be a solution of the
Einstein equations.

To prove this we write the gyraton metric \Eq{AdS.gyraton} in the
form
\ba\n{13}
g_{\alpha\beta}=\bar{g}_{\alpha\beta}+2l_{(\alpha} a_{\beta)},
\ea
where $\bar{g}_{\alpha\beta}$ is the AdS metric \Eq{AdS}
and the vectors $l_{\alpha}$ and $a_{\alpha}$ (in the coordinates
adopted in (\ref{AdS}) and (\ref{AdS.gyraton}))  have the components
\ba
l_u&=&-{L^2\over z^2},\hskip 0.5cm l_v=l_a=0,\\
a_u&=&{1\over 2}\Phi,\hskip 0.5cm a_v=0,\hskip 0.5cm
a_a=A_a(u,{\bf x}).
\ea
In what follows we shall use the gyraton metric $g_{\alpha\beta}$ to
operate with indices. For example, $a^\alpha=g^{\alpha\beta}a_\beta$
and $l^\alpha=g^{\alpha\beta}l_\beta$. In particular, one has
\be
l^{\alpha}=\delta^{\alpha}_{v}\hh
l^{\alpha}a_{\alpha}=0\, .
\ee
It is easy to see that ${\bf l}=l^{\alpha}\pa_{\alpha}$ is the Killing
vector in the both metrics (\ref{AdS}) and (\ref{AdS.gyraton}). This vector
obey the following properties
\ba\n{ll}
l^\epsilon l_\epsilon&=&0,\hskip 0.5cm 
l^\epsilon_{;\epsilon}=0, \\ \n{lK}
l_{\alpha;\beta}&=&2l_{[\alpha}\kappa_{\beta]} \hh l^\epsilon
\kappa_\epsilon=0 ,
\ea
where $\kappa_\alpha=-\nabla_\alpha\ln z$. Here semicolon denotes the covariant
derivative with respect to the gyraton metric. We use notations
$l_{(\alpha}a_{\beta)}={1\over
2}\left(l_{\alpha}a_{\beta}+l_{\beta}a_{\alpha}\right)$ and 
$l_{[\alpha}\kappa_{\beta]}={1\over
2}\left(l_{\alpha}\kappa_{\beta}-l_{\beta}\kappa_{\alpha}\right)$.

Straightforward calculations show that the scalar curvature for the
metric \Eq{AdS.gyraton} is constant $R=-{D(D-1)\over L^2}$. The
Riemann tensor for the gyraton metric can be written as the sum 
\ba\n{Rr}
R_{\alpha\beta\mu\nu}=\hat{R}_{\alpha\beta\mu\nu}+r_{\alpha\beta\mu\nu}.
\ea
Here the constant curvature part reads
\ba\n{hatR}
\hat{R}_{\alpha\beta\mu\nu}
=\left[g_{\alpha\mu}g_{\beta\nu}-g_{\alpha\nu}g_{\beta\mu}\right]R.
\ea

Let us express the Riemann tensor of the AdS metric 
$\bar{g}_{\alpha\beta}$ in terms of the gyraton metric $g_{\alpha\beta}$.
We have
\ba\n{barg}
\bar{g}_{\alpha\beta}=g_{\alpha\beta}-2l_{(\alpha}a_{\beta)}.
\ea
The inverse AdS metric, being expressed in terms of the gyraton
metric, has a form
\ba\n{bargup}
\bar{g}^{\alpha\beta}=g^{\alpha\beta}+2l^{(\alpha}a^{\beta)}
+l^{\alpha}l^{\beta}a^{\epsilon}a_{\epsilon}.
\ea
Straightforward calculation of the Christoffel symbols gives
\ba\n{gamm}
\bar{\Gamma}^\mu_{\alpha\beta}=\Gamma^\mu_{\alpha\beta}-\gamma^\mu_{\alpha\beta},
\ea
where 
\ba\nn
\gamma^\mu_{\alpha\beta}&=&l^\mu a_{(\alpha;\beta)}
+2 a_{(\alpha} l^\mu_{;\beta)}+l_{(\alpha}F_{\beta)}{}^\mu\\ 
&+&2l^\mu a^\epsilon l_{\epsilon;(\alpha}a_{\beta)}
+l^\mu l_{(\alpha}F_{\beta)\epsilon}a^\epsilon. \n{ga}
\ea
Here $F_{\alpha\beta}=\pa_{\alpha} a_{\beta}-\pa_{\beta} a_{\alpha}$.
The `spatial' components of this tensor, $F_{ab}$, coincide with
(\ref{Fab}), and $F_{\alpha\beta}l^{\beta}=0$.
It is easy to check that the tensor $\gamma^\mu_{\alpha\beta}$ obeys
the relations
\ba\nn
\gamma^\mu_{\alpha\beta}l_\mu=\gamma^\mu_{\alpha\beta}l^\alpha
=\gamma^\mu_{\alpha\beta}l^\beta=0.
\ea
Using (\ref{ga}) and the property \Eq{lK} one can show that
\ba\n{gamma-pq}
\gamma^\mu_{\alpha\beta}=l^\mu
p_{\alpha\beta}+l_{(\alpha}q_{\beta)}{}^{\mu},
\ea
\ba
l^{\alpha}p_{\alpha\beta}=l^{\beta}q_{\beta}{}^{\mu}=l_{\mu}q_{\beta}{}^{\mu}=0.
\ea
We call a tensor to be aligned to the the vector $l_{\alpha}$  if it
can be written as a sum of terms, where each term contains as
a factor at  least one vector $l_{\alpha}$. 
The tensor $\gamma^\mu_{\alpha\beta}$ is aligned to the vector
$l_{\alpha}$.

Substituting the decomposition \Eq{gamm} into the definition of the
Riemann tensor
\ba\n{RR}
R^\mu{}_{\nu\alpha\beta}=\partial_{\alpha}\Gamma^\mu_{\nu\beta}
-\partial_{\beta}\Gamma^\mu_{\nu\alpha}
+\Gamma^\mu_{\epsilon\alpha}\Gamma^\epsilon_{\nu\beta}
-\Gamma^\mu_{\epsilon\beta}\Gamma^\epsilon_{\nu\alpha}
\ea
we obtain
\ba\nn
\bar{R}^\mu{}_{\nu\alpha\beta}=R^\mu{}_{\nu\alpha\beta}
-\nabla_{\alpha}\gamma^\mu_{\nu\beta}+\nabla_{\beta}\gamma^\mu_{\nu\alpha}
+\gamma^\mu_{\epsilon\alpha}\gamma^\epsilon_{\nu\beta}
-\gamma^\mu_{\epsilon\beta}\gamma^\epsilon_{\nu\alpha}.
\ea
It's easy to see that all terms containing $\gamma^\mu_{\nu\beta}$
and $\nabla_{\alpha}\gamma^\mu_{\nu\beta}$ are aligned with
$l_\alpha$  owing to \Eq{gamma-pq} and \Eq{lK}. The difference of the
Riemann tensors with all indices in the lower position has the same
property because lowing index on the left hand-side with \Eq{barg}
results to additional term $-l^\mu
a_{\epsilon}R^\epsilon{}_{\nu\alpha\beta}$ on the right hand-side,
which is orthogonal to $l_\alpha$. Therefore, the difference
$r_{\mu\nu\alpha\beta}$ of the Riemann tensors  
\ba\nn
\bar{R}_{\mu\nu\alpha\beta}=R_{\mu\nu\alpha\beta}-r_{\mu\nu\alpha\beta}
\ea
is aligned with the vector $l_\alpha$ and has the form
\ba\n{r0}
r_{\mu\nu\alpha\beta}=l_{[\mu}
K_{\nu][\alpha\beta]}+l_{[\alpha}K_{\beta][\mu\nu]},
\ea
where the tensor $K_{\mu[\alpha\beta]}$ is orthogonal to $\l^\alpha$
\ba\nn
l^{\mu}K_{\mu[\alpha\beta]}=l^{\alpha}K_{\mu[\alpha\beta]}=0.
\ea


\subsection{Curvature invariants}

At first we consider scalar invariants constructed from the curvature but
which do not contain covariant derivatives. Scalar invariants,
constructed  from powers of the curvature  $R_{\alpha\beta\mu\nu}$
involve the powers of $\hat{R}_{\alpha\beta\mu\nu}$, \Eq{hatR},
constructed from the gyraton metric, and the powers of
$r_{\alpha\beta\mu\nu}$. But the powers of $r_{\alpha\beta\mu\nu}$
contain at least one $l_\alpha$ which must be contracted either with
another $l_\alpha$,  or with $a_\alpha$ and $K_{\alpha\mu\nu}$. In
all these cases the contraction gives zero. Therefore, only the
powers of $\hat{R}_{\alpha\beta\mu\nu}$ may survive a contraction
over all indexes. Hence, such invariants calculated for the gyraton
metric are identical to those of the exact AdS geometry.

Now consider scalar invariants containing covariant derivatives of
the curvature with respect to the gyraton metric $g_{\alpha\beta}$.
The derivatives of the constant curvature part in the decomposition
\Eq{Rr} vanish identically.  So, there remain only terms with
derivatives acting on  $r_{\alpha\beta\mu\nu}$ given by \Eq{r0}.
Since $r_{\alpha\beta\mu\nu}$ is aligned with the vector ${\bf l}$,
terms which enter a given scalar invariant always contain either
${\bf l}$ or its covariant derivatives. Besides these terms, the
scalar invariant may also include terms constructed from  
$a_\alpha$, $\kappa_\alpha$, $K_{\alpha\mu\nu}$ and their covariant
derivatives. In the general case the scalar invariant is a linear
combination of products of such terms. For each of the product one can
define the total number of covariant derivatives which enter the
product. Suppose that $n$ is the largest of these numbers, than we
denote the corresponding scalar invariant by $S_n$. We call $n$ an
order of the invariant. 

In what follows we shall use the property that ${\bf l}$ is the
Killing vector. Denote by $T_{\alpha \ldots}^{ \beta \ldots}$ a
tensor constructed from $a_\alpha$, $\kappa_\alpha$,
$K_{\alpha\mu\nu}$ and their covariant derivatives. Then
\be
{\cal L}_{\bf l} T_{\alpha \ldots}^{ \beta \ldots}=0\, ,
\ee
or using the definition of the Lie derivative 
\be\n{lie}
l^{\mu}\nabla_{\mu}  T_{\alpha \ldots}^{ \beta \ldots}=
\nabla_{\mu} l^{\alpha} \ T_{\mu \ldots}^{ \beta \ldots}+\ldots 
-
\nabla_{\beta} l^{\mu} \ T_{\alpha \ldots}^{ \mu \ldots}-\ldots
\, .
\ee

Consider a scalar invariant $S_n$. We describe now operations which
allows one to transform identically $S_n$ into a `canonical'  form.
Each of these operations either keeps the order $n$ the same or reduces
it. 

{\em Operation 1}. If in an invariant $S_n$ there exists a covariant
derivative acting in ${\bf l}$, one can use \Eq{lK} to exclude it.
By repeating this procedure one can always transform $S_n$ into the
form $S'_{n'}$ without derivatives of ${\bf l}$. It is evident that
$n'\le n$.

{\em Operation 2}. Consider an invariant $S_n$ which does not contain
derivatives of ${\bf l}$. Since ${\bf l}^2=0$, the vector index of
$l^{\alpha}$ must be contracted either with the index of $a_\alpha$,
$\kappa_\alpha$, $K_{\alpha\mu\nu}$, or with one of the covariant
derivatives. In the former case, the corresponding term is of the
form $l^{\alpha} \nabla \ldots \nabla p_{\ldots \alpha \ldots}$,
where $p_{\ldots \alpha \ldots}$ is one of the tensors $a_\alpha$,
$\kappa_\alpha$, $K_{\alpha\mu\nu}$. Using the relation
\be\n{com}
l_{\alpha} \nabla_{\beta} (\ldots)=\nabla_{\beta}\left[ l_{\alpha}
(\ldots)\right] - 2l_{[\alpha}\kappa_{\beta ]} (\ldots)\, ,
\ee
one can `move' ${\bf l}$ through covariant derivatives. By repeating
this procedure and using \Eq{lK} one finally tranform $S_n$ into
the form
\be
S_n=\nabla \ldots \nabla (l^{\alpha} p_{\ldots \alpha
\ldots})+S_{n-1}\, .
\ee
Since $l^{\alpha} p_{\ldots \alpha \ldots} =0$ as a final result of
this procedure one can exclude terms containing $l^{\alpha} \nabla
\ldots \nabla p_{\ldots \alpha \ldots}$ and decrease the order $n$. By
using again (if necessary)  the Operation 1 one can
transform the obtained expression $S_{n-1}$ into the form without the
derivatives of ${\bf l}$.

{\em Operation 3}. Consider now a term where the index of ${\bf l}$ is
contracted with the index of a covariant derivative
\be
l^{\alpha} \ldots \nabla_{\alpha} (\ldots)\, .
\ee
Using (\ref{com}) one can transform the corresponding invariant $S_n$
(without increasing its  order) into a form 
\be
\ldots l^{\alpha} \nabla_{\alpha} (\ldots)\, .
\ee
By using  (\ref{lie}) and \Eq{lK} one can decrease the number of
derivative in this expression. As the result, one reduces $S_n$ to
$S_{n-1}$. By using (if necessary)  the
Operation 1 one can transform the obtained expression $S_{n-1}$ into
the form without the derivatives of ${\bf l}$.

By repeating the operations 2 and 3 one finally arrives to an
invariant of the zero order in derivatives. But all zero order
invariants necessarily contain as factors the contraction of
$l^\epsilon$ with $a_\epsilon$, $\kappa_\epsilon$, and 
$K_{\epsilon\mu\nu}$, which are zero. Thus, only invariants which
did not contain derivatives may not vanish. But, as was shown the
beginning of this subsection, these invariants for AdS and
AdS-gyraton metrics are identical.  Therefore, all scalar invariants
for the gyraton in the asymptotically AdS spacetime and scalar
invariants of exact AdS spacetime coincide. It should be emphasized
that we do not use  the fact that $\Lambda$ is negative. Thus the
same result is also valid in the asymptotically de Sitter spacetime.


\section{Scalar Green function in AdS}\label{AppendixA}
The Euclidean propagator for a massless scalar field is the
solution to the equation
\ba\nn
\bBox G(x,x')=-\delta(x,x'), \hskip 0.5cm x^A=(z,x^4,\dots,x^{N+2}).
\ea
Here we enumerated coordinates starting with $x^3=z$, rather than $x^1$
in order to distinguish this artificial Euclidean AdS from the
physical spacetime.
The Euclidean $AdS_N$ metric with a unit radius $L=1$ reads
\ba\n{EuclideanAdS}
ds^2={1\over z^2}\left[(dz)^2+(dx^4)^2+\dots+(dx^{N+2})^2 \right].
\ea
The geodesic distance $\mu(x,x')$ in this geometry is
\ba\nn
\cosh(\mu(x,x'))=1+U(x,x'),
\ea
\ba\n{Ugeneral}
U={\delta_{AB}(x^A-x'^A)(x^B-x'^B)\over 2zz'}.
\ea
It obeys the equation
\ba
\mu_A\mu^A=1,\hskip 1cm U_A U^A=U(U+2).
\ea
Because AdS is highly symmetrical space all necessary physical
quantities may be expressed in terms of only one biscalar function
(see, e.g., \cite{AllenJacob:1986})
$\mu(x,x')$ or $U\equiv U_N(x,x')$. Thus for the Green function in  even
$(N=2k)$ and odd $(N=2k+1)$ number of dimensions we obtain
accordingly \cite{DanKesKru:1999}
\ba\nn
G_{2k}(x,x')&=&-{1\over (2\pi)^k}\left(-{\partial\over\partial
U}\right)^{k-1}Q_{k-1}\left(U+1\right),\\ \n{G2k}
G_{2k+1}(x,x') &=&-{1\over (2\pi)^k}\left(-{\partial\over\partial
U}\right)^{k-1}\\ \nn
&\times&{\left[U+1-\sqrt{U(U+2)}\right]^k\over 2\sqrt{U(U+2)}}.
\ea
Here $Q_{k-1}(U+1)$ is the Legendre function and $U=U_N(x,x')$.

For the problem in question the source on the right hand-side  of the
\Eq{ADSscalar} depends on the retarded time $u$ and
$(D-2)$ spatial coordinates $(z,x^4,\dots,x^{D})$. Therefore, the
solution for the scalar potential $\Phi$ will require integration
over all unphysical extra dimensions. For our purpose we need to
know an integral of the scalar Green function over two extra
coordinates (e.g., $x^{N+1}$ and $x^{N+2}$).
\ba\nn
\int dx^{N+1} dx^{N+2} G(U_N)=2\pi zz'\int_{U_{N-2}}^\infty dU_N~G(U_N).
\ea
This property trivially follows from the relation
\ba\nn
U_N&=&U_{N-2}+{l^2\over 2zz'}, \\
\nn
l^2&=&(x^{N+1}-x'^{N+1})^2+(x^{N+2}-x'^{N+2})^2.
\ea
The Legendre function of integer index can be expressed in terms of
elementary functions. For example explicit expressions for the Green
functions from 2 till 6 dimensions are
\ba\nn
G_2&=&{1\over 4\pi}\ln\left({U\over U+2}\right),
\\ \nn
G_3&=&-{1\over 4\pi}\left[{U+1\over\sqrt{U(U+2)}}-1\right],
\\ 
\n{scalarG}
G_4&=&-{1\over 8\pi^2}\left[\ln\left({U\over U+2}\right)+{1\over
U}+{1\over U+2}\right],
\\ \nn
G_5&=&-{1\over
8\pi^2}\left[2+{1-3U-6U^2-2U^3\over(U(U+2))^{3/2}}\right],
\\ \nn
G_6&=&{1\over 16\pi^3}\left[3\ln\left({U\over U+2}\right)
+{2(U+1)(3U^2+6U-2)\over U^2(U+2)^2}\right].
\ea


\section{Green function for a massless vector field in AdS}\label{AppendixB}

Consider a Green function $G_{AB'}(x,x')$ for massless vector fields
in Euclidean AdS space. This propagator satisfies the equation
\ba
\nabla^A\nabla_{[A}G_{B]B'}=-g_{BB'}\delta(x,x')
+\partial_{B'}\Lambda_B(x,x')
\ea
and can be represented in the form
\ba\n{Gmunu}
G_{AB'}(x,x')=-(\partial_{A}\partial_{B'}U)~F(U)+\partial_{A}\partial_{B'}S(U).
\ea
Here $F(U)$ describes the propagation of the physical components of
$A_A$ while $\Lambda_B$ and $S(U)$ are gauge artifacts and can be discarded
\cite{AllenJacob:1986,HFMMR:1999}.
\ba\nn
F(U)=C_N~[U(2+U)]^{1-N/2},
\hskip 0.7cm C_N={\Gamma\left({N\over2}-1\right)\over(4\pi)^{N/2}}.
\ea
\ba\nn
\partial_{A}\partial_{B'}U&=&-{1\over zz'}\left[
\delta_{AB'}-U\delta_{A z}\delta_{zB'}\right.\\
&& \nn \left. +{1\over z}(x-x')_A\delta_{zB'}
-{1\over z'}(x-x')_{B'}\delta_{A z}.
\right].
\ea
The vector potential is defined as
\ba
A_A(x)=\int d^Nx'\sqrt{g'}G_{AB'}(x,x')J^{B'}(x').
\ea
Similar to the scalar field case
the current $J^{B'}$ doesn't depend on some coordinates $x^k$ and doesn't
have components in these directions. Then integration over these
coordinates reduces to the integration of the scalar function  $F(U)$.

Integral of $F(U)$ over two "redundant" coordinates gives
\ban
\int dx^{N+1} dx^{N+2}~F(U_N)=2\pi zz'\int^\infty_{U_{N-2}}dU_N~F(U_N),
\ean
\ban
\int^\infty_{U_{N-2}}dU_N~F(U_N)&=&C_N{(U_{N-2})^{3-N}\over N-3}\\
&\times&F(N-3,{N\over 2}-1;N-2;-{2\over U_{N-2}}).
\ean
For example, for $N=4$
\ban
{1\over C_4}\int^\infty_{U_{2}}dU_4~F(U_4)
=-{1\over 2}\ln{U_2\over U_2+2};
\ean
for $N=5$
\ban
{1\over C_5}\int^\infty_{U_{3}}dU_5~F(U_5)
={U_3+1\over \sqrt{U_3(U_3+2)}}-1;
\ean
for $N=6$
\ban
{1\over C_6}\int^\infty_{U_{4}}dU_6~F(U_6)
={1\over 4}\left[\ln{U_4\over U_4+2}+{1\over U_4}+{1\over U_4+2}\right];
\ean
for $N=7$
\ban
{1\over C_7}\int^\infty_{U_{5}}dU_7~F(U_7)
&=&-{2\over 3}{U_5+1\over \sqrt{U_5(U_5+2)}} \\
&&+{2\over 3}
+{1\over 3}{U_5+1\over [U_5(U_5+2)]^{3\over2}}.
\ean
The integral of the function $F(U)$ over four "redundant" coordinates gives
\ban
&&\int dx^{N-1} dx^{N} dx^{N+1} dx^{N+2}F(U_N)\\
&&=(2\pi zz')^2 \int^\infty_{U_{N-4}}dU_{N-2}\int^\infty_{U_{N-2}}dU_N~F(U_N)\\
&&=C_N{(2\pi zz')^2 U_{N-4}^{4-N}\over (N-3)(N-4)}~F(N-4,{N\over
2}-1;N-2;-{2\over U_{N-4}}).
\ean
For example, one has for $N=5$
\ban
{1\over C_5}\int^\infty_{U_1}dU_{1}\int^\infty_{U_3}dU_3~F(U_5)
=-\sqrt{U_1(U_1+2)}+U+1;
\ean
for $N=6$
\ban
{1\over C_6}\int^\infty_{U_2}dU_{4}\int^\infty_{U_4}dU_6~F(U_6)
={1\over 4}(U_2+1)\ln\left({U_2\over U_2+2}\right)+{1\over 2};
\ean
for $N=7$
\ban
{1\over C_7}\int^\infty_{U_3}dU_{5}\int^\infty_{U_5}dU_7~F(U_7)
&=&{2\over 3}\left[
\sqrt{U_3(U_3+2)}-U-1\right.\\
&&\left. +{1\over\sqrt{U_3(U_3+2)}}\right].
\ean



\begin{thebibliography}{}

\bibitem{To} R. C. Tolman, {\em Relativity, Thermodynamics, and
Cosmology}, p.272, Oxford, Clarendon Press (1934).

\bibitem{Pe1} A. Peres, Phys. Rev. Lett. {\bf 3}, 571 (1959).

\bibitem{Pe2} A. Peres, Phys. Rev.  {\bf 118}, 1105 (1960).

\bibitem{Bo} W. B. Bonnor, Commun. Math. Phys. {13}, 163 (1969).

\bibitem{FrFu:05} V.P. Frolov and D.V. Fursaev,
Phys.Rev. {\bf D71}, 104034 (2005); hep-th/0504027.

\bibitem{FroIsrZel:05}
V.P. Frolov, W. Israel, and A. Zelnikov, hep-th/0506001.

\bibitem{AiSe} P.C. Aichelburg and R.U. Sexl,
Gen.Rel.Grav. {\bf 2}  303 (1971).

\bibitem{Siklos} S.T.C. Siklos, in {\it Galaxies, Axisymmetric Systems and
Relativity}, edited by M.A.H. MacCallum (Cambridge University Press, Cambridge,
England, 1985);\\ 
J. Podolsky, Class.Quant.Grav. {\bf 15} 719 (1998).

\bibitem{DrayTHooft:85} T. Dray and G.'t Hooft,
Nucl.Phys.B {\bf 253}  173 (1985);
Class.Quant.Grav.{\bf 3}  825 (1986).

\bibitem{Ferrari:88} V. Ferrari, P. Pendenza, G. Veneziano,
Gen.Rel.Grav. {\bf 20}  1185 (1988);
H. de Vega, N. Sanchez, Nucl.Phys.B {\bf 317}  706 (1989).

\bibitem{Barrabes:01} C. Barrabes, P.A. Hogan
Phys.Rev. D {\bf 64}  0044022 (2001).

\bibitem{Sfetsos:95} K. Sfetsos, Nucl.Phys.B {\bf 436}  721 (1995).

\bibitem{Kaloper:05} N. Kaloper,
Phys.Rev. {\bf D71}  086003 (2005); hep-th/0502035.

\bibitem{HorowitzSteif:90}
G.T. Horowitz, A.R. Steif, Phys.Rev.Lett. {\bf 64}  260 (1990).

\bibitem{AmatiKlimcik:89}
D. Amati, C. Klimcik, Phys.Lett. {\bf B219}  443 (1989).

\bibitem{HorowitzItzhaki:99} G.T. Horowitz, N. Itzhaki,
{\bf JHEP} 9902:010 (1999); hep-th/9901012.

\bibitem{Brinkmann:25}
H.W. Brinkmann,
Math.Ann. {\bf 94}  119 (1925).

\bibitem{Tseytlin:95} A.A. Tseytlin,
Class.Quant.Grav. {\bf 12}  2365 (1995); \\
M. Blau, M. O`Loughlin, G. Papadopoulos and A.A. Tseytlin,
Nucl.Phys.B {\bf 673}  57 (2003).

\bibitem{Podolsky:98} J. Podolsky,
Class.Quant.Grav. {\bf 15} 3229 (1998).

\bibitem{Kallosh:98}
R. Kallosh, A. Rajaraman, Phys.Rev. {\bf D58}  125003 (1998); hep-th/9805041.
 
\bibitem{AllenJacob:1986}
B. Allen, T. Jacobson, Commun.Math.Phys. {\bf 103} 669 (1986).

\bibitem{DanKesKru:1999}
U.H. Danielsson, E. Keski-Vakkuri and M. Kruczenski, JHEP {\bf 9901} 002
(1999); hep-th/9812007.

\bibitem{HFMMR:1999}
E. D'Hoker, D.Z. Freedman, S.D. Mathur, A. Matusis, L. Rastelli, Nucl.Phys.B
{\bf 562} 330 (1999); hep-th/9902042.

\bibitem{HorowitzHubeny:00} g.t. Horowitz, V.E. Hubeny,
{\bf JHEP} 0010:027 (2000); hep-th/0009051.

\end{thebibliography}
\end{document}